\documentclass[useAMS,usenatbib,letters]{mn2e}

\usepackage{amsmath}
\usepackage{amssymb}
\usepackage{graphicx}
\usepackage{xcolor}
\usepackage{url}
\usepackage{natbib}

\title[The puzzle of the CNO abundances of $\alpha$ Cygni variables]{The puzzle of the CNO abundances of $\alpha$ Cygni variables resolved by the Ledoux criterion?}
\author[C. Georgy et al.]{Cyril Georgy$^{1}$\thanks{Email: c.georgy@keele.ac.uk}, Hideyuki Saio$^{2}$, Georges Meynet$^{3}$\\
$^1$ Astrophysics, Lennard-Jones Laboratories, EPSAM, Keele University, Staffordshire, ST5 5BG, UK \\
$^2$ Astronomical Institute, Graduate School of Science, Tohoku University, Sendai, Japan\\
$^3$ Geneva Observatory, University of Geneva, Maillettes 51, 1290 Versoix, Switzerland}

\begin{document}

\maketitle

\begin{abstract}
Recent stellar evolution computations show that the blue supergiant (BSG) stars could come from two distinct populations: a first group arising from massive stars that just left the main sequence (MS) and are crossing the Hertzsprung-Russell diagram (HRD) towards the red supergiant (RSG) branch, and a second group coming from stars that have lost considerable amount of mass during the RSG stage and are crossing the HRD for a second time towards the blue region. Due to very different luminosity-to-mass ratio, only stars from the second group are expected to have excited pulsations observable at the surface. In a previous work, we have shown that our models were able to reproduce the pulsational properties of BSGs. However, these models failed to reproduce the surface chemical composition of stars evolving back from a RSG phase. In this paper, we show how the use of the Ledoux criterion instead of the Schwarzschild one for convection allows to significantly improve the agreement with the observed chemical composition, while keeping the agreement with the pulsation periods. This gives some support to the Ledoux criterion.
\end{abstract}


\section{Introduction}

The evolution of massive stars after the main sequence (MS) remains mostly unknown, despite numerous improvements of stellar evolution codes during the last twenty years. Actually, it strongly depends on several physical processes that are suspected to take place in stellar interiors, such as internal mixing, rotation, magnetic fields, of which the exact implementation in evolution codes is still mostly uncertain. Mass loss also plays a key role, particularly during the red supergiant (RSG) phase, but is also a major source of uncertainties \citep{vanLoon2005a,vanBeveren1998a,vanBeveren1998b,Georgy2012a}.

On the observational side, the determination of the maximal luminosity of Galactic RSGs \citep{Levesque2005a} at around $\log(L/L_{\sun}) = 5.6$ (bolometric magnitude of $\sim -9.2$) and recent stellar evolution models \citep{Ekstrom2012a} indicate that the most luminous RSGs should originate from stars up to $\sim 25-30\,M_{\sun}$. On the other hand, \citet{Smartt2009a} determined that the maximal mass for a type IIP supernova (SN) progenitor is around $17\,M_{\sun}$. As RSGs are expected to explode as type IIP SNe, there is here an indication that at least the most massive RSGs do not end their life as RSGs, but evolve further to become stars of other types.

The recent release by the Geneva group of a new set of rotating models at solar metallicity \citep{Ekstrom2012a} shows that these massive RSGs can cross the Hertzsprung-Russell diagram (HRD) a second time towards the blue side, if the mass-loss rates used during the RSG phase are increased by a factor of 3\footnote{\footnotesize{These increased rates are still compatible with the rates inferred by \citet{vanLoon2005a}.}}. Stars with an initial mass above $20\,M_{\sun}$ explode as luminous blue variables or Wolf-Rayet stars \citep{Georgy2012b,Groh2013a,Groh2013b}.

In this framework, two distinct populations of blue supergiant (BSG) stars are expected: the first one (group 1 hereafter) composed by the stars leaving the MS and going to the RSG branch (first crossing of the HRD), and the second one (group 2 hereafter) by the stars that were on the RSG branch, and that cross a second time the HRD for any reason. In a previous paper \citep{Saio2013a}, we showed that these two populations have very distinct pulsational properties, allowing to distinguish them. Group 1 stars exhibit few excited pulsation modes at the surface, due to their low $L/M$ ratio. On the contrary, stars of the group 2 have a lot of excited modes, that are compatible with the observed pulsation period of $\alpha$-Cyg type variable stars. However, at least two stars (Rigel and Deneb) for which the period is compatible with our predicted period for group 2 have chemical surface abundances \citep{Przybilla2010a} that are compatible with our stellar evolution models for group 1, but not for group 2.

In this paper, we show how to tackle this discrepancy by using another prescription for the convection in our stellar models. In Section~\ref{Sec_Models}, we describe the evolution of our new models, and explain how they solve the problem of the chemical abundances. In Section~\ref{Sec_Pulsation}, we show that these new models are still compatible with the pulsational properties of observed BSG stars. Finally, our conclusions are presented in Section~\ref{Sec_Conclu}.

\section{Evolution of the surface abundances}\label{Sec_Models}

\subsection{The Schwarzschild model}\label{SubSec_Schwarzschild}

\begin{figure*}
\begin{center}
\includegraphics[width=.45\textwidth]{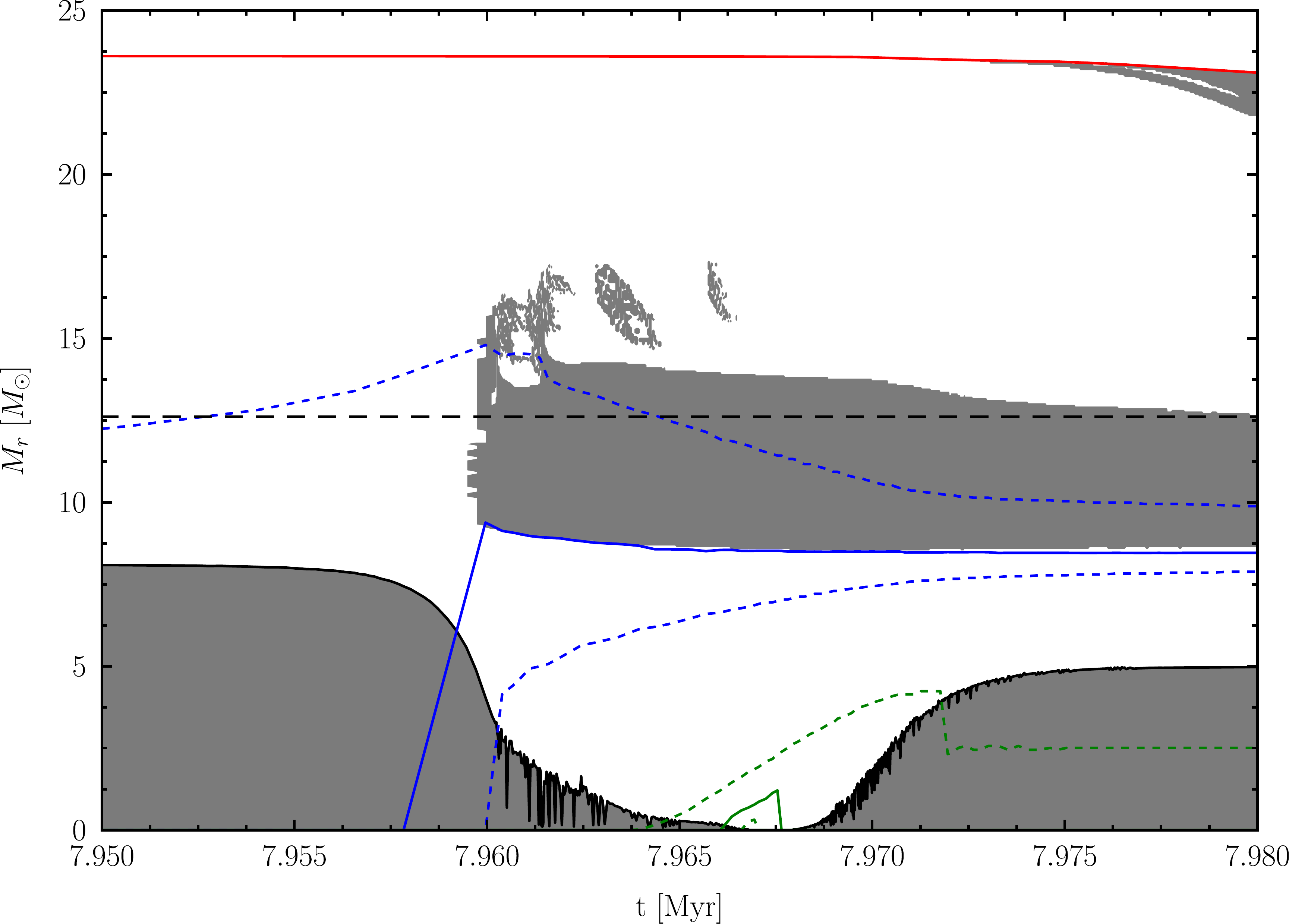}\hfill\includegraphics[width=.45\textwidth]{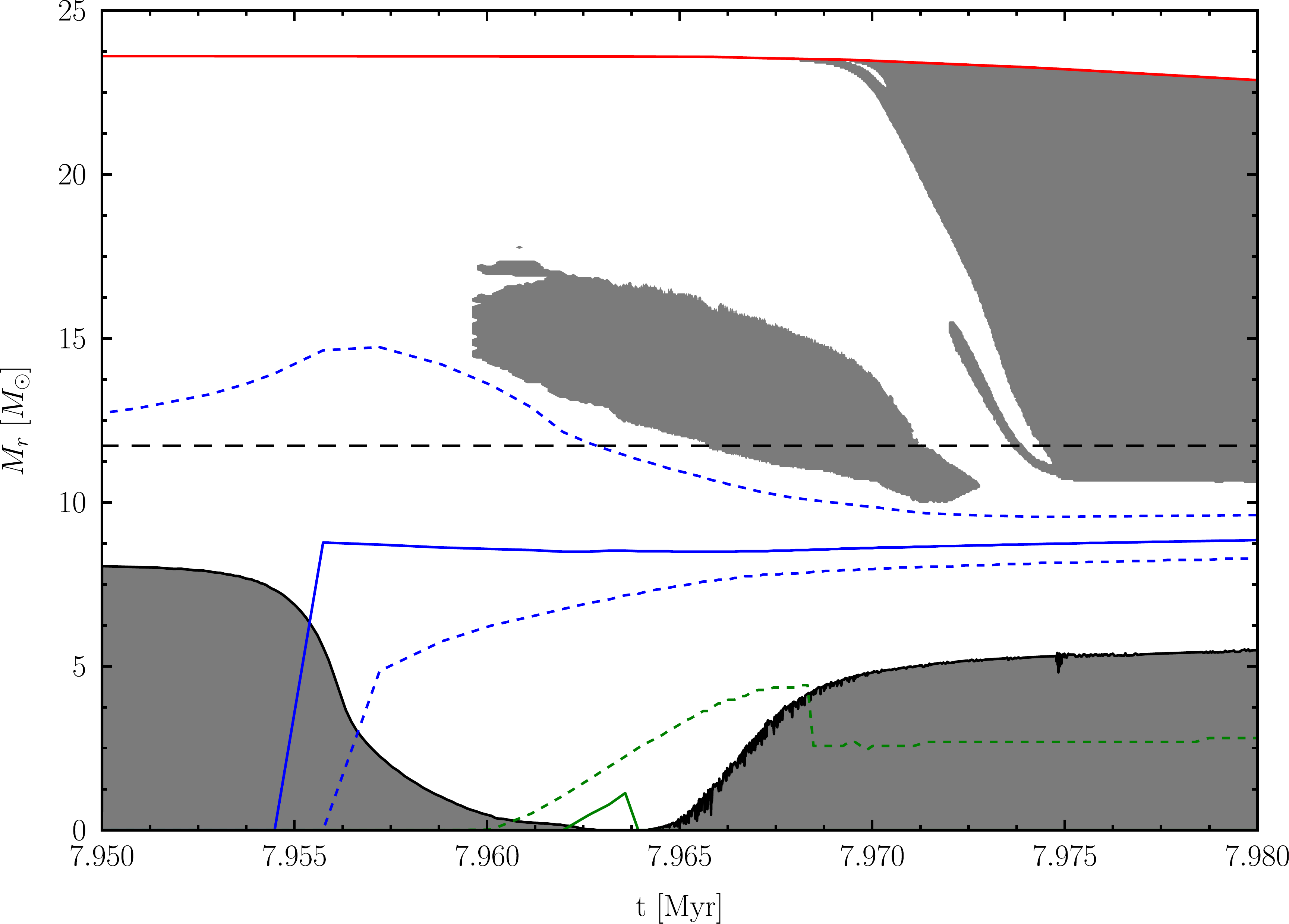}
\end{center}
\caption{Kippenhahn diagram of a rotating $25\,M_{\sun}$, from the very end of the MS up to the appearance of the external convective zone. The red line indicates the position of the surface, and the grey zones represents the convective regions. The blue (respectively green) lines indicate the position where H- (respectively He-) burning energy generation rate is maximal (solid line), and $100\,\text{erg}\cdot\text{s}^{-1}\cdot\text{g}^{-1}$ (dotted lines). The dashed black line indicates the layer that is uncovered (due to mass loss) when the star has $\log(T_\text{eff}) = 4.0$. \textit{Left panel:} Model computed with the Schwarzschild criterion. \textit{Right panel:} Model computed with the Ledoux criterion.}
\label{Fig_Kipp}
\end{figure*}

In this Section, we discuss the main processes affecting the evolution of the chemical abundances at the surface of the rotating $25\,M_{\sun}$ model we showed in \citet{Saio2013a}. This will serve as a basis for the discussion of the new model presented below. The physical ingredients used to compute this model are exactly the same as in \citet{Ekstrom2012a}. Particularly, the Schwarzschild criterion for convection was used. In addition, an overshoot distance equal to $0.1H_P$ beyond the Schwarzschild boundary was considered, where $H_\text{P}$ is the pressure scale height.

After the MS, the characteristic evolution timescale becomes short enough to neglect the rotational mixing in the discussion\footnote{\footnotesize{At least at solar metallicity and for the masses considered here. The situations is different, for example, at low metallicity \citep{Meynet2002a}.}}. The chemical structure of the star is thus mostly affected by the various nuclear-burning episodes (in the centre or in shells), and by the development of convective zones. The surface abundances are also affected by the mass loss.

As shown in Fig.~\ref{Fig_Kipp} (left panel), two phenomena contributes to the change of the surface chemical composition of the surface after the MS: the development of a convective zone appearing when hydrogen burning migrates from the centre to a shell (at $t=7.96\,\text{Myr}$ in the figure), and the development of an external convective zone when the star reaches the coldest part of the HRD ($\log(T_\text{eff}) \lesssim 3.6$, at $t=7.975\,\text{Myr}$ in the figure). The intermediate convective zone plays an important role, as it homogenises a region that will be later uncovered by the strong mass loss that these stars encounter during the RSG phase. The total mass of the star when it reaches $\log(T_\text{eff}) = 4.0$ during the second crossing of the HRD is indicated by the dashed black line in Fig.~\ref{Fig_Kipp}.

Figure~\ref{Fig_ChemSch} shows the chemical structure of the star at different times. At the end of the MS (top-left panel), we have a central region completely depleted in H, and with the typical imprint of CNO burning: increased N abundance, and depleted C and O. Going towards the surface, we have a chemical gradient progressively bringing the abundances to almost the ZAMS values (in this rotating model, the rotational mixing has slightly changed the surface composition due to the diffusion of the chemical species).

Just after the MS and before central He ignition, an intermediate H-burning convective shell appears immediately on top of the core (the shaded area in the top-right panel of Fig.~\ref{Fig_ChemSch}). The effect of this deep convective zone is the creation of a quite extended region strongly depleted in C and O, and enriched in N, between roughly $10$ and $15\,M_{\sun}$.

At the arrival on the RSG branch (bottom left panel), the situation is almost unchanged, except for the appearance of a quite small external convective zone below the surface, that homogenises the chemical composition. During the later evolution of the star, the huge mass loss encountered during the RSG phase progressively uncovers the layer below the surface. When the star reaches $\log(T_\text{eff})=4.0$ (bottom-right panel), the uncovered layers are those that laid previously in the H-burning convective shell, with a strong N enrichment, and C and O depletion. As we showed in \citet{Saio2013a}, this composition is not compatible with the observed one at the surface of at least 2 BSGs (Rigel and Deneb).

\subsection{Model with the Ledoux criterion}

Trying to reconcile the pulsational properties of our models during the second HRD crossing (that are compatible with the observations of variable BSGs) with the observed surface chemical composition (that are compatible with models during the first crossing), the following options are available:
\begin{itemize}
\item Changing the mass-loss rates.
\item Changing the internal mixing processes.
\end{itemize}

\begin{figure}
\begin{center}
\includegraphics[width=.23\textwidth]{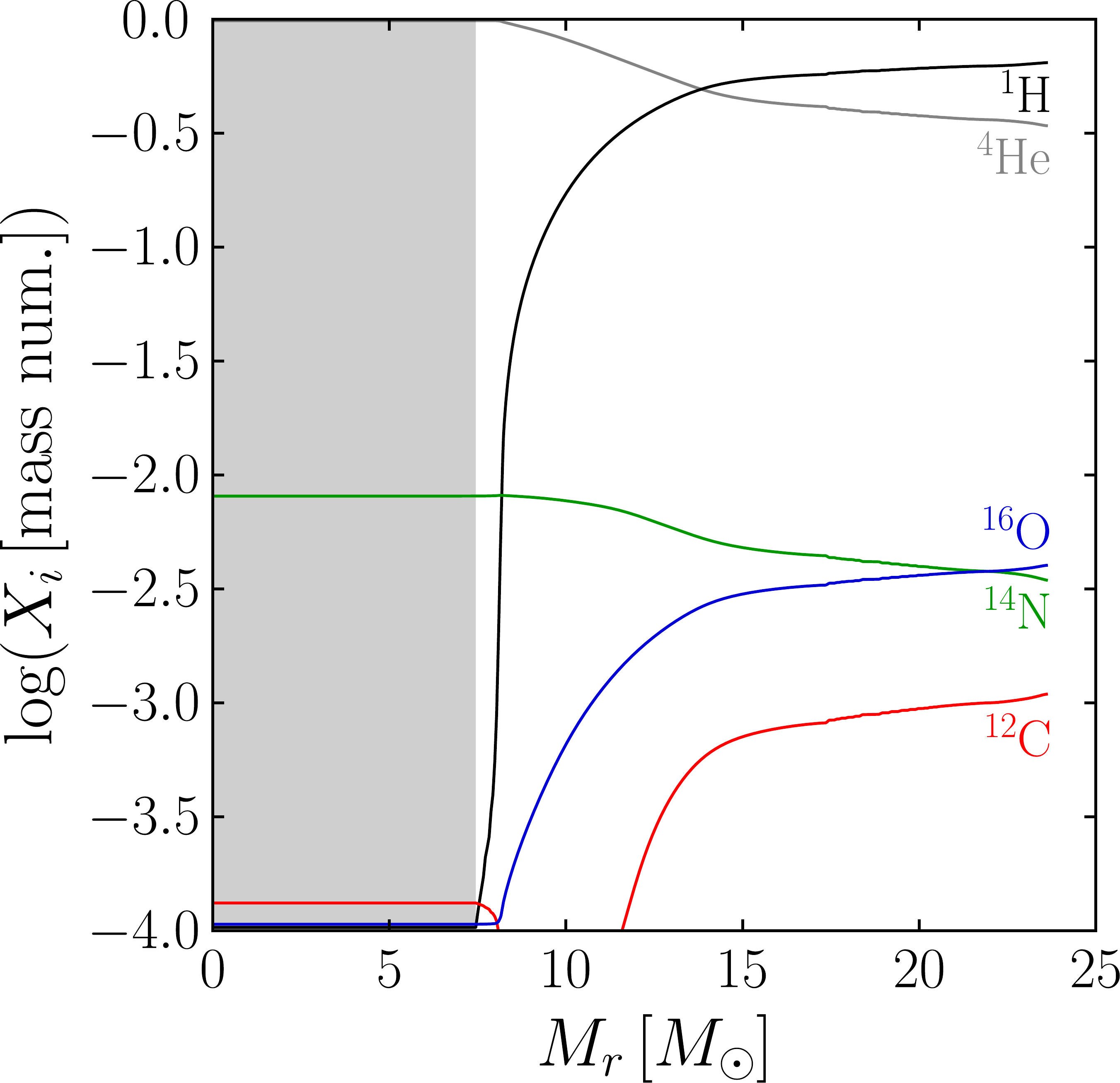}\hfill\includegraphics[width=.23\textwidth]{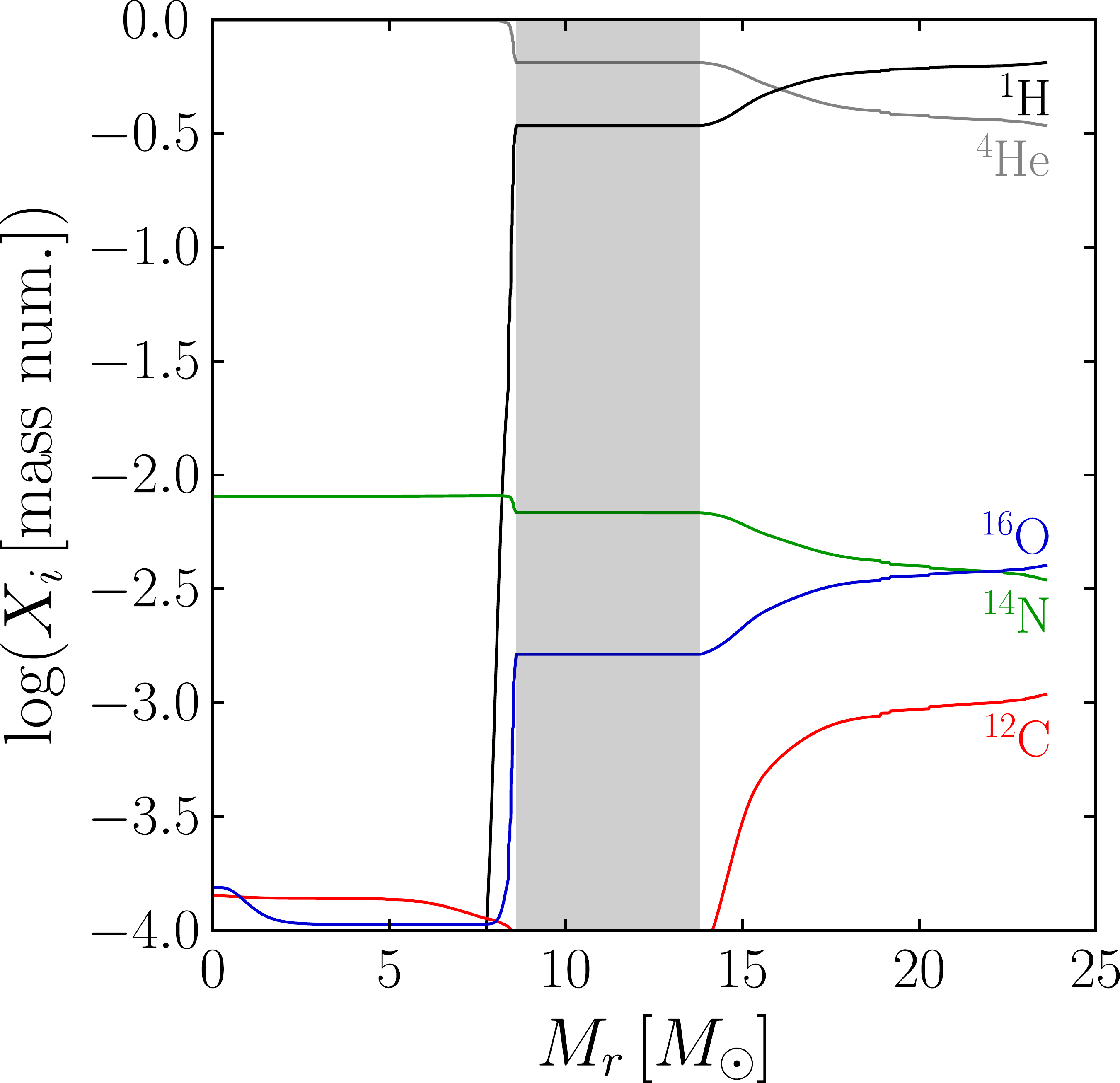}\\
\includegraphics[width=.23\textwidth]{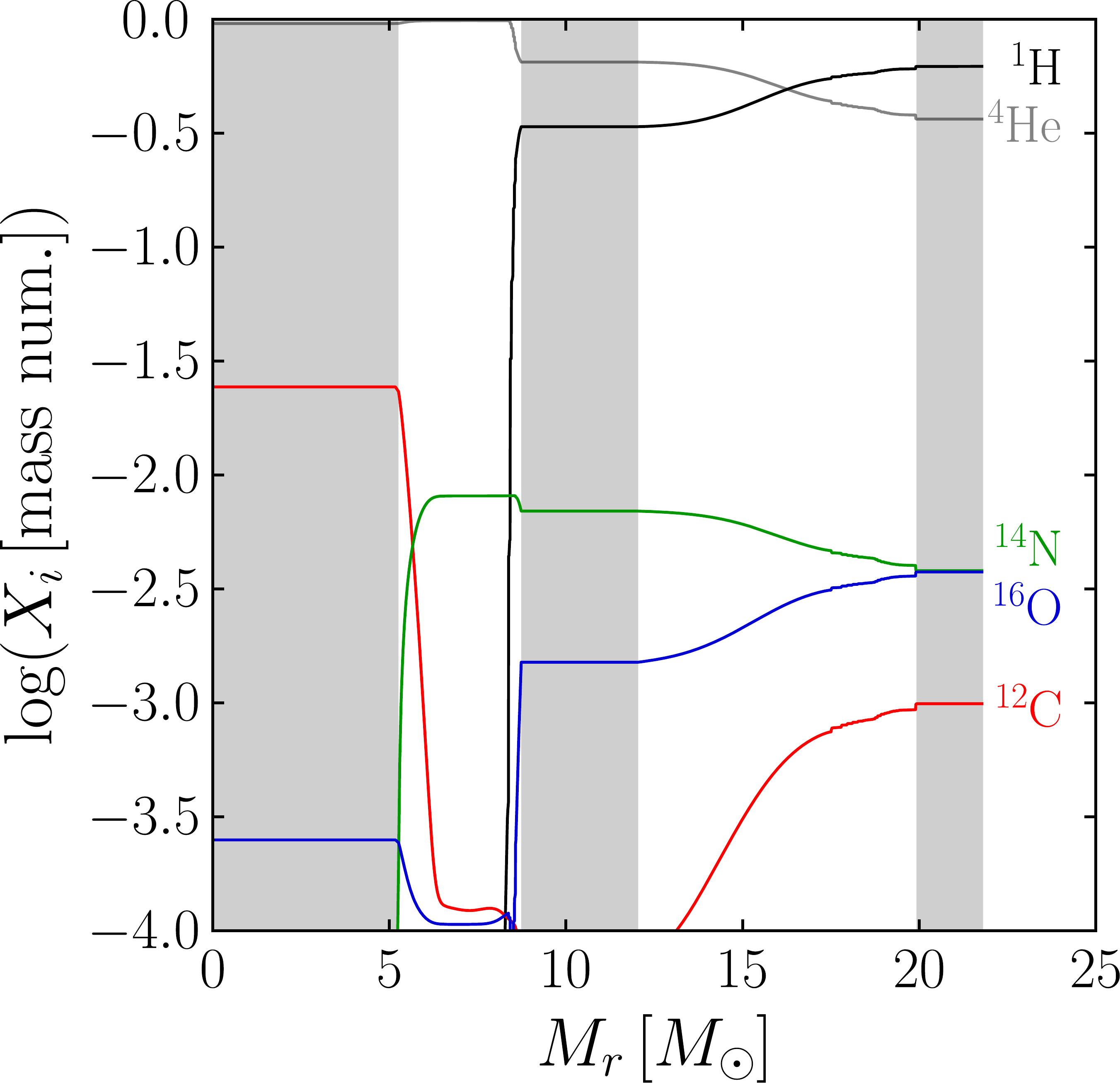}\hfill\includegraphics[width=.23\textwidth]{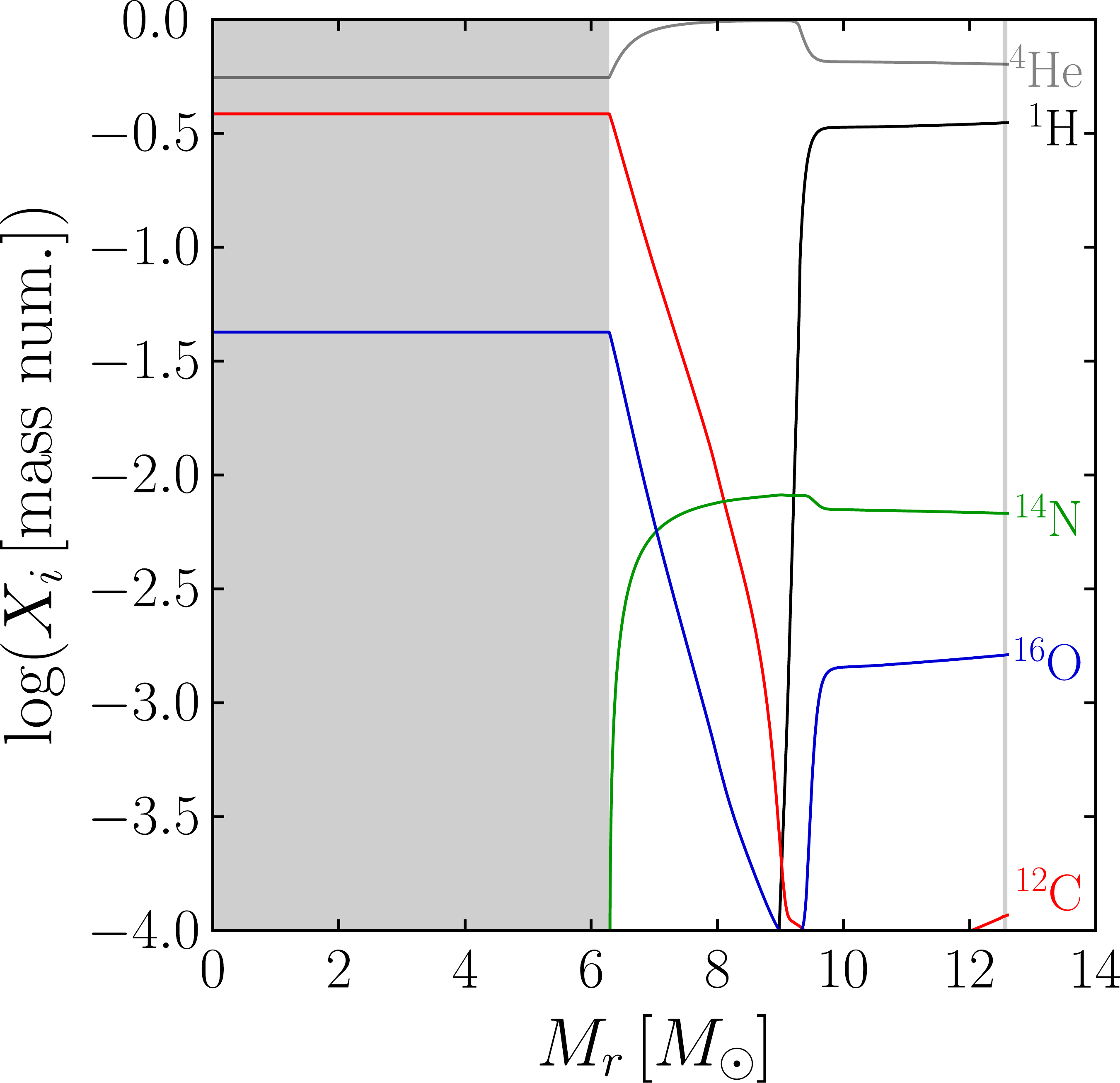}
\end{center}
\caption{Chemical structure of the rotating $25\,M_{\sun}$ computed with the Schwarzschild criterion for convection. The logarithms of the abundances of H (black), He (grey), C (red), N (green), and O (blue, all in mass fraction) is plotted as a function of the Lagrangian mass coordinate (the star's centre is on the left, the surface on the right). The shaded areas represent the convective regions. \textit{Top left panel:} End of the MS; \textit{top right panel:} He-core burning ignition and H-shell burning; \textit{bottom left panel:} First arrival on the RSG branch; \textit{bottom right panel:} Second HRD crossing when the star reaches $\log(T_\text{eff}) = 4$.}
\label{Fig_ChemSch}
\end{figure}

\begin{figure}
\begin{center}
\includegraphics[width=.23\textwidth]{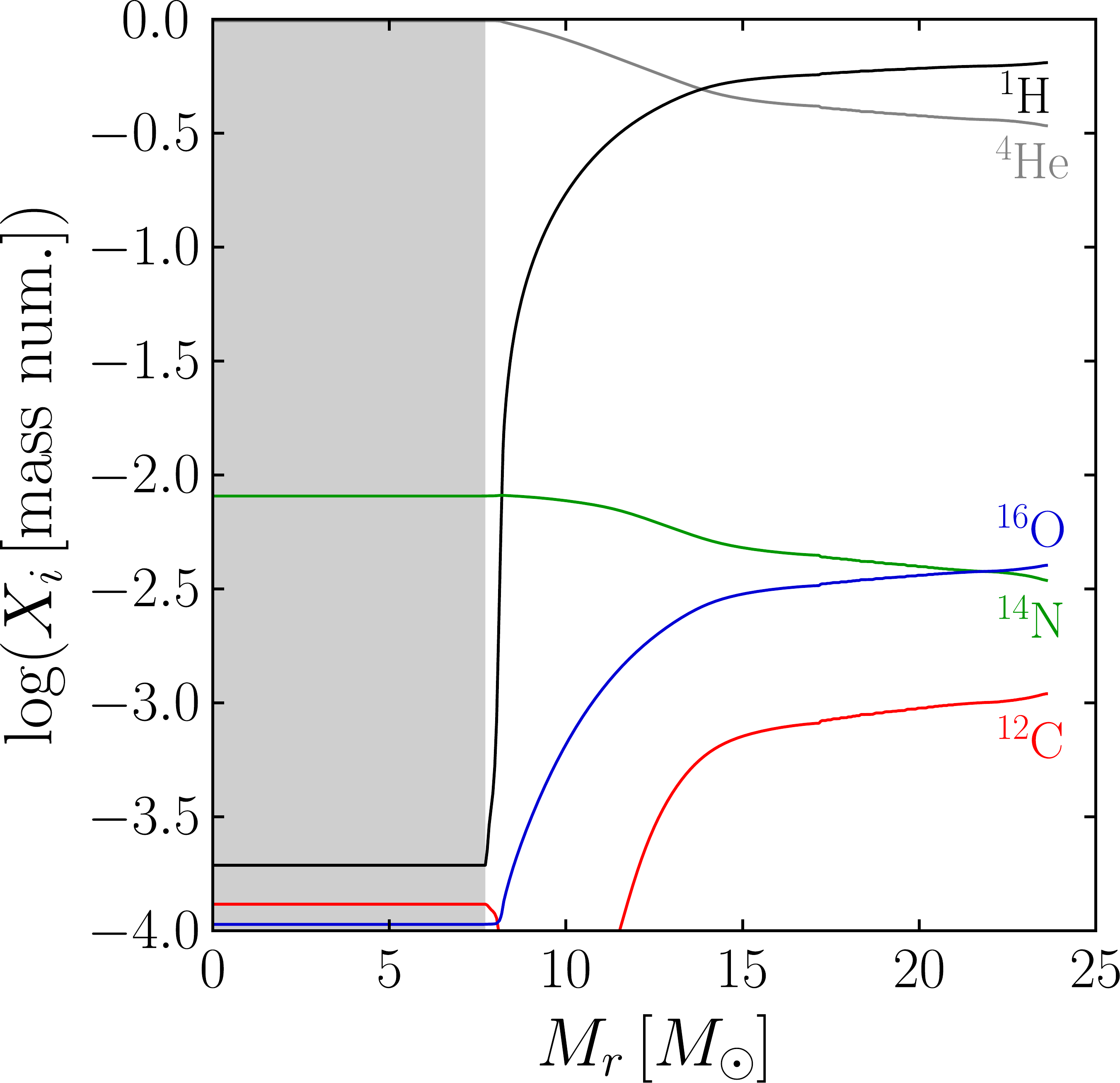}\hfill\includegraphics[width=.23\textwidth]{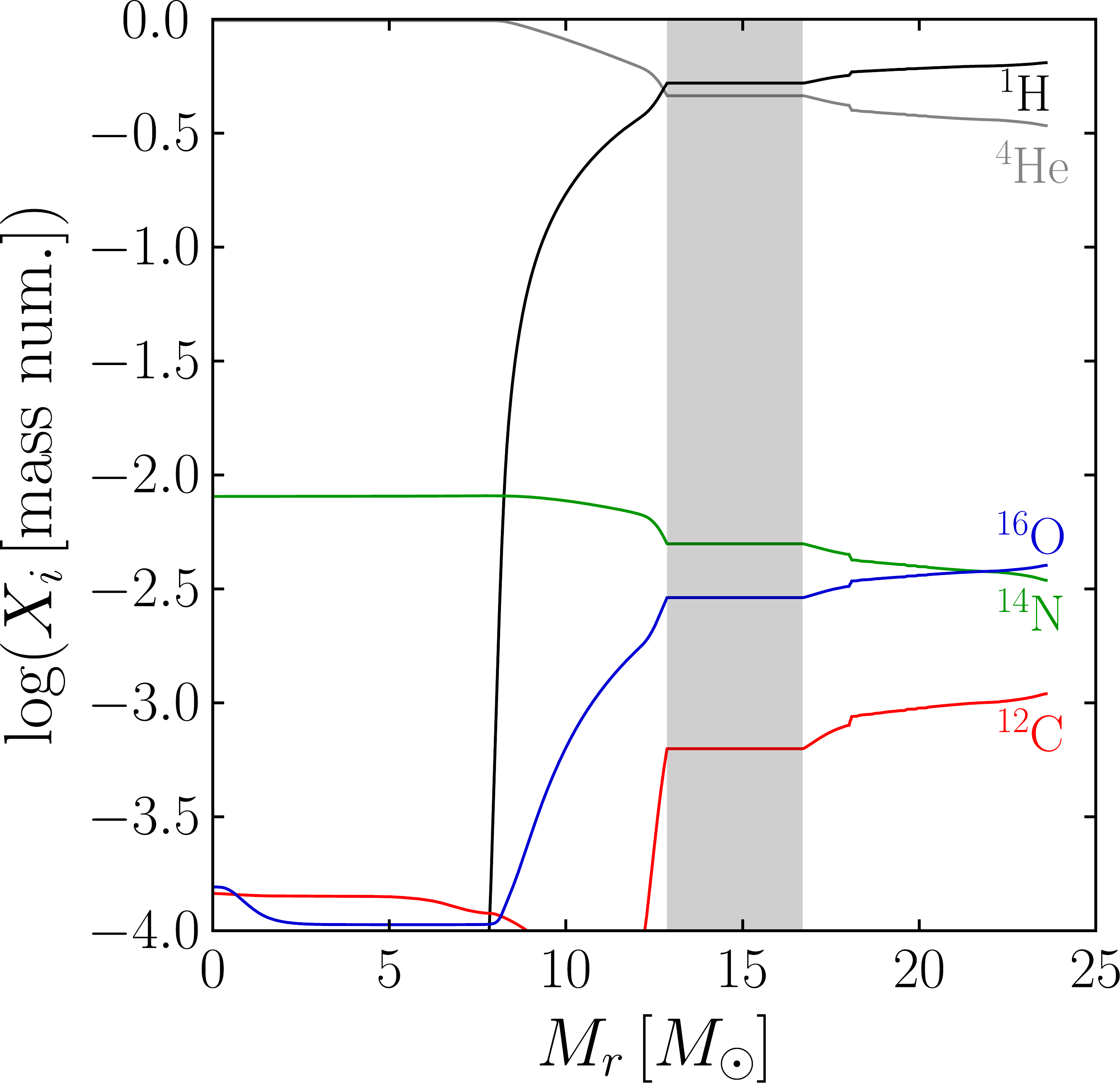}\\
\includegraphics[width=.23\textwidth]{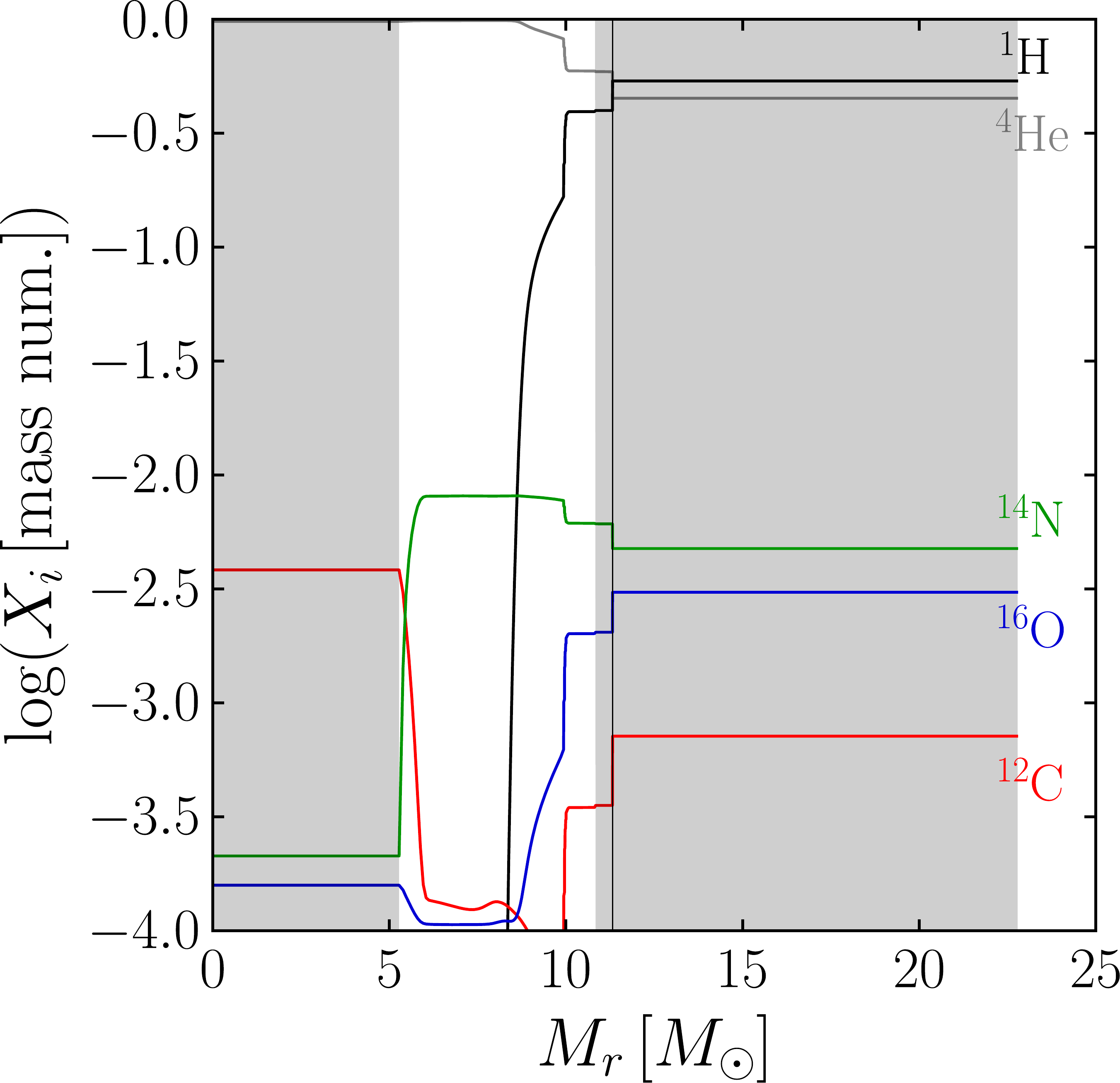}\hfill\includegraphics[width=.23\textwidth]{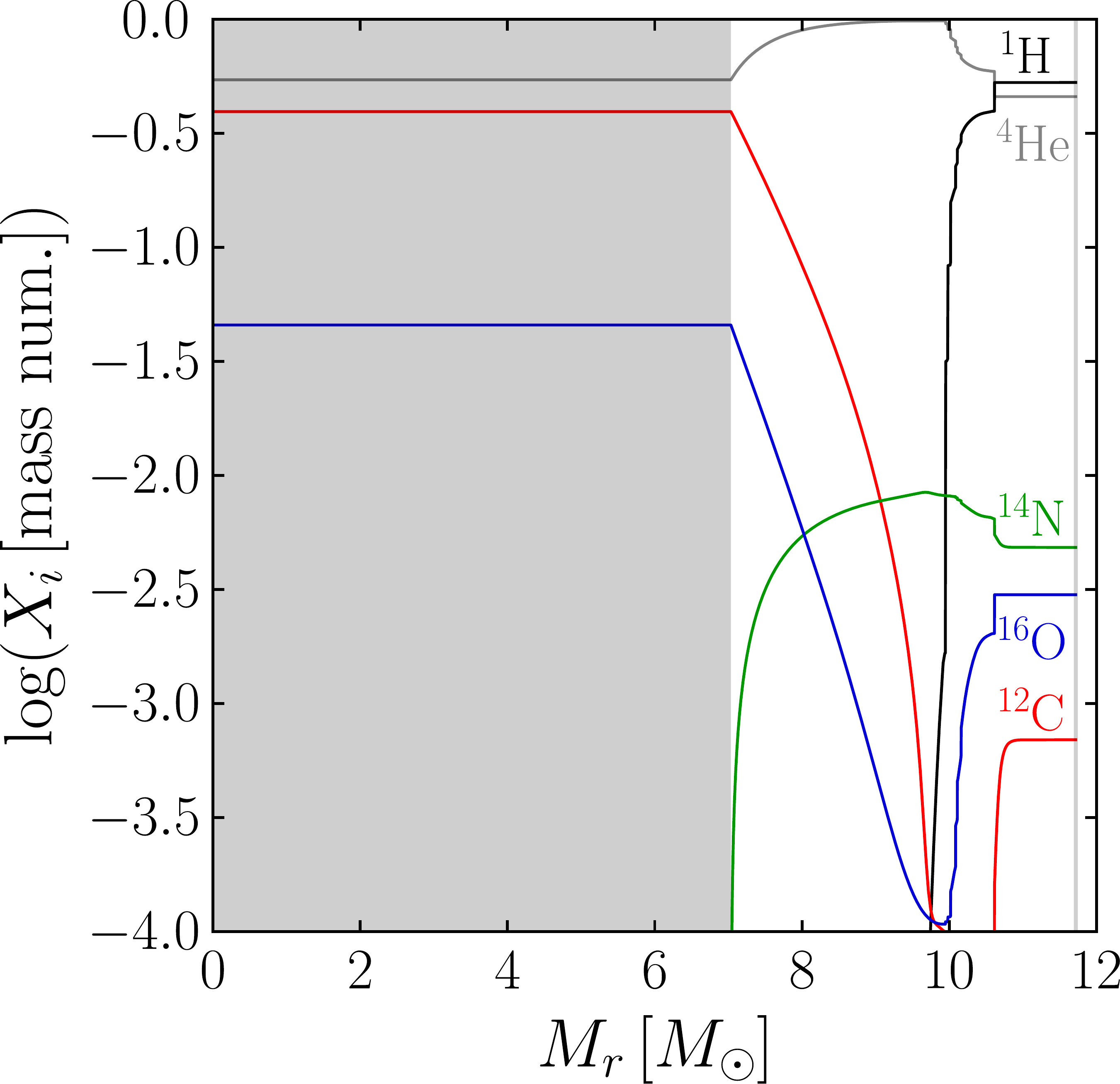}
\end{center}
\caption{Same as Fig.~\ref{Fig_ChemSch}, but for the model computed with the Ledoux criterion.}
\label{Fig_ChemLed}
\end{figure}

Keeping exactly the same physical properties, but varying the mass-loss rates during the RSG phase does not allow to significantly change the surface chemical composition of BSGs during the second crossing. It changes the time spent on the RSG branch before the second crossing occurs, but the surface chemical composition when the star reaches the BSG region for the second time is roughly independent of the mass-loss rates used. Actually, for the star to evolve blue wards from the RSG stage, a fixed amount of mass has to be lost. However, the timescale for this loss does not really matter. Indeed, what determines the surface composition of the BSG in group 2 is the composition of the layers that are exposed to the surface due to this mass removal and the composition of these layers is already fixed (at least in the two models shown in Figs~\ref{Fig_ChemSch} and \ref{Fig_ChemLed}) at the entrance into the RSG stage.

To change the chemical structure of the star at the first entrance into the RSG stage, we have to modify the way the elements are mixed in the interior. On the one hand, we can change the mixing in the radiative zones, using other prescription for the rotational mixing \citep[see][]{Meynet2013a}. Some tests were performed in that direction without obtaining significant changes in the surface abundances of BSGs in the group 2 with respect to the computation detailed in Section~\ref{SubSec_Schwarzschild}. This can be understood since rotational mixing is the most efficient during the MS (due to the long timescale available), and as the chemical abundances during the second crossing depends on phenomena that occur \textit{after} the MS.

On the other hand, we can change the treatment of convection in our models. In our recent works \citep{Ekstrom2012a,Georgy2013a,Georgy2013b}, we used the Schwarzschild criterion to determine the convective region. However, another criterion commonly used in the literature is the Ledoux one \citep[e.g.][]{Brott2011a}. To examine the impact of changing this criterion in our model, we computed a model strictly equivalent to the one presented in Section~\ref{SubSec_Schwarzschild}, only changing the criterion for convection. In particular, note the following: 1) We kept the same overshooting as in the Schwarzschild case, in order that our rotating model is still able to reproduce the width of the MS for lower mass stars \citep[see][]{Ekstrom2012a}. In addition, with the instantaneous mixing approximation for convective mixing done in the Geneva code, this prevents any differences to develop between the convective core sizes computed with the Ledoux and Schwarzschild criteria. 2) No semi-convective mixing is applied in our models. As semi-convection develops in region that are Ledoux stable, but Schwarzschild unstable, any model accounting for this effect should behaves between the two set of models presented in this paper: a very inefficient semi-convection corresponding to our ``Ledoux'' model, and a very efficient semi-convection to our ``Schwarzschild'' model.

Figure~\ref{Fig_Kipp} (right panel) shows the evolution of the stellar structure at the very end of the MS, up to the reaching of the RSG branch for the ``Ledoux'' model. Comparing with left panel, we notice the following trends:
\begin{itemize}
\item There is no obvious difference between the size of the core in both models.
\item In the ``Schwarzschild'' model, the intermediate convective zone is linked with the development of the H-burning shell. In the ``Ledoux'' model, the convective zone is inhibited in the deepest region due to the strong $\mu$-gradient at the edge of the core. Since the H-shell burning is not linked with a convective zone, it is weaker than the H-burning shell of the ``Schwarzschild'' model, and disappears quickly. This makes the star to cross the HRD towards the RSG branch much more rapidly in the ``Ledoux'' case, associated with a decrease of the luminosity. This allows the appearance of a deep external convective zone.
\end{itemize}

In Fig.~\ref{Fig_ChemLed} we show the internal chemical structure of the star, at the same evolutionary stages as in Fig.~\ref{Fig_ChemSch}. At the end of the MS, the structure between both models are the same. Divergences in the behaviour appears immediately after central H exhaustion, when the H-burning migrates in a shell. In the ``Schwarzschild'' case, it is associated with a deep intermediate convective region, producing a quite extended zone in the star strongly affected by the CNO cycle (see Sect.~\ref{SubSec_Schwarzschild}). Since the external convective zone is relatively small, the region in the star where the chemical abundances are compatible with the measurements at the surface of Deneb and Rigel (see Table~\ref{Tab_Chem}) is close to the surface (between a Lagrangian mass coordinate $M_r\sim 18-24$, see top-right panel of Fig.~\ref{Fig_ChemSch}). In order to evolve toward the blue, the star has to uncover much deeper layers (deeper than about $6\,M_{\sun}$), which makes the surface composition incompatible with the one observed at the surface of Rigel and Deneb.

On the contrary, in the case of the ``Ledoux'' model, the intermediate convective zone is less deep, keeping the zone with a strong C and O depletion deeper in the star compared to the ``Schwarzschild'' case (top-right panel of Fig.~\ref{Fig_ChemLed}). Moreover, the development of a deep external convective zone on the RSG branch (bottom-left panel) will strongly dilute the CNO products. These two factors imply that the BSGs in group 2 will present much weaker N/C and N/O ratios as can be seen in Table~\ref{Tab_Chem}. Adopting the Ledoux criterion instead of the Schwarzschild one strongly decreases the N/C and N/O ratios for the BSGs in the group 2. Here we did not try to obtain a perfect match with the observed values for Rigel and Deneb. However, adopting a lower initial rotation would likely allow to obtain a still better agreement. This is not the case when the Schwarzschild criterion is used because the structure at the entrance in  the RSG phase is fundamentally different whatever the rotation rate.

\begin{table}
\caption{Surface chemical ratio in the ``Schwarzschild'' and ``Ledoux'' models of a rotating $25\,M_{\sun}$ when the surface $\log(T_\text{eff}) = 4.0$ during the second crossing of the HRD, as well as the observed values for Rigel and Deneb \citep{Przybilla2010a}.}
\label{Tab_Chem}
\begin{center}
\begin{tabular}{cccc}
model/star & $\text{N}/\text{C}$ & $\text{N}/\text{O}$ & $X_\text{He}$\\
\hline
``Schwarzschild'' (model) & $57.86$ & $4.17$ & $0.635$ \\
``Ledoux'' (model) & $6.97$ & $1.61$ & $0.458$ \\
Rigel (observation) & $2.0$ & $0.46$ & $0.32$ \\
Deneb (observation) & $3.4$ & $0.65$ & $0.37$
\end{tabular}
\end{center}
\end{table}

\section{Pulsational properties}\label{Sec_Pulsation}

\begin{figure}
\begin{center}
\includegraphics[width=.41\textwidth]{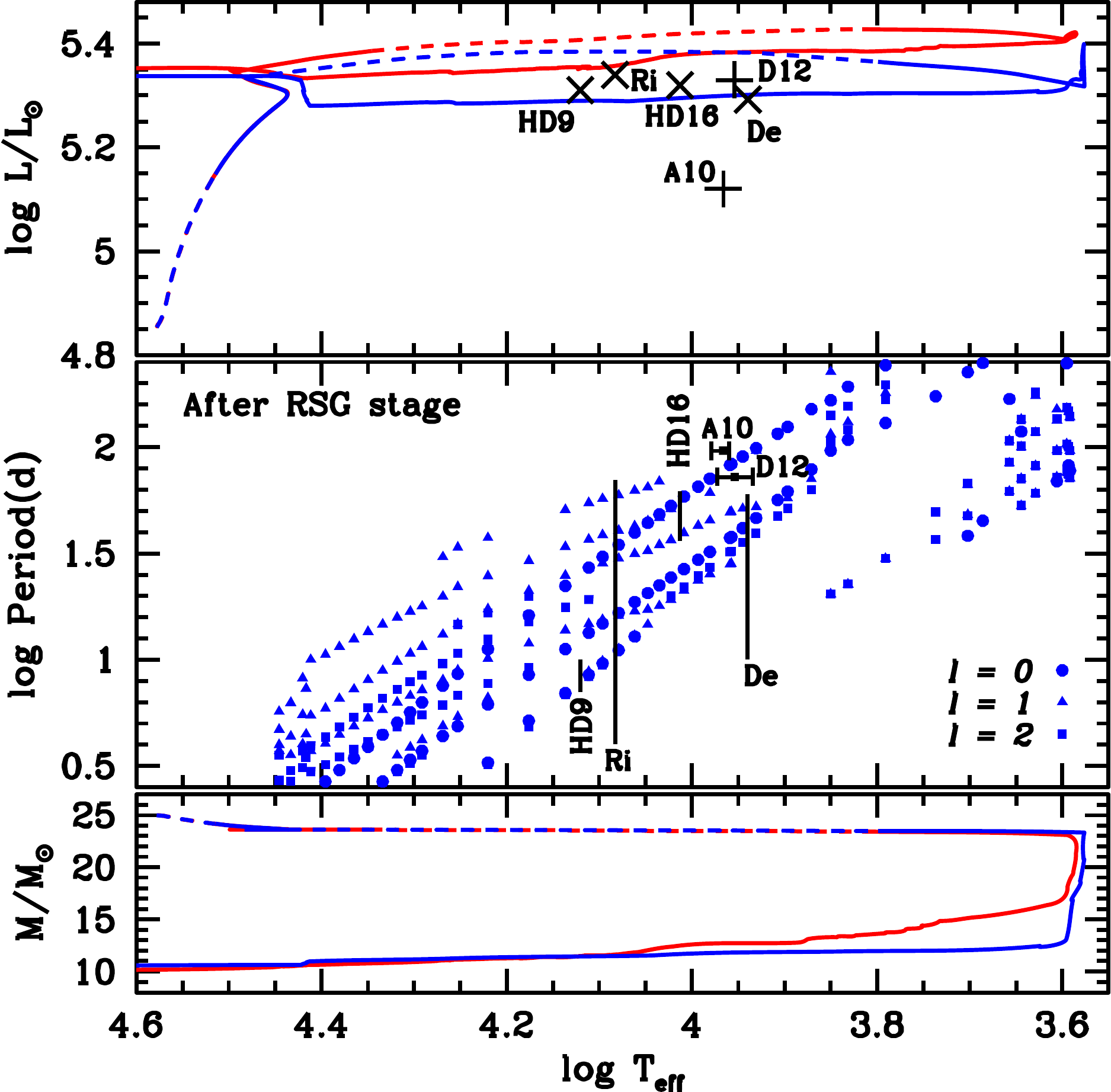}
\end{center}
\caption{\textit{Top panel:} HRD of the $25\,M_{\sun}$ computed with the Ledoux (blue) and the Schwarzschild (red) criteria for convection. The solid line indicates models in which radial pulsations are excited, while dashed line indicates models for which pulsation stability is examined but no radial modes are excited. \textit{Middle panel:} Logarithm of the pulsation period as a function of the effective temperature for the model with Ledoux criterion, as well as the observed period range for the same stars as in top panel. The periods (or period ranges) are shown for some $\alpha$ Cyg variables in the Milky Way (crosses) and in NGC 300 (pluses) that have luminosities within the considered range. The references of the observational data and the names of the stars are given in \citet{Saio2013a}. \textit{Bottom panel:} Total mass of  as a function of $T_\text{eff}$.}
\label{Fig_HRDPuls}
\end{figure}

Fig.~\ref{Fig_HRDPuls} shows evolutionary tracks (upper panel) and the evolution of the total mass (lower panel) for models with the Schwarzschild criterion (red) and the Ledoux criterion (blue) for convection. Evolutionary tracks in the MS are nearly independent of the convection criterions, while the post-MS model with the Ledoux criterion is generally less luminous.
  
Pulsation properties are insensitive to the convection criterions as long as the models lie at the same position in the HRD and have the same mass-luminosity ratio. In both cases,  during the first crossing after  the  MS, no radial pulsations are excited in the range $3.8 \lesssim \log(T_\text{eff}) \lesssim 4.3$, where most of the $\alpha$ Cygni variables are located, while radial pulsations are excited during the second crossing, due to the larger $L/M$ ratio produced by the strong mass loss during the RSG phase (see bottom panel of Fig.~\ref{Fig_HRDPuls}).

We compare observed periods with those predicted for $25\,M_{\sun}$ models with the Ledoux criterion in the second crossing in Fig.~\ref{Fig_HRDPuls}. Many radial and non-radial pulsations are excited in these models; the excitation mechanisms and the amplitude distributions in the stellar interior are discussed in detail in \citet{Saio2013a}. 

Also shown are observational properties of some $\alpha$ Cygni variables whose luminosities are comparable to  the evolutionary track. Most of the $\alpha$ Cygni type variations are quasi periodic, which indicates that several periods are simultaneously excited. Such period ranges are represented by vertical lines in the middle panel of Fig.~\ref{Fig_HRDPuls}. The variable A10 and D12 in NGC 300 are rare exceptions; \citet{Bresolin2004a} obtained regular light curves with periods of $96.1$ and $72.5$ days, respectively, indicating that their pulsation is dominantly a single radial mode.

As we can see in this figure, the periods and period ranges of $\alpha$ Cygni variables are more or less explained by the pulsations excited in the second crossing models. Only the shorter period range of Deneb cannot be explained by our models. This difficulty is also present for the models with the Schwarzschild criterion as discussed in \citet{Saio2013a}.   

We found in this section that the pulsational properties of the $25\,M_{\sun}$ models with the Ledoux criterion in the second crossing are similar to those with the Schwarzschild criterion, and they largely agree with the properties of $\alpha$ Cygni variables having  comparable luminosities. This confirms that models computed with the Ledoux criterion provides in our case better agreement with the observations than the Schwarzschild criterion.

\section{Conclusions}\label{Sec_Conclu}

In this paper, we have studied how the choice of the criterion for convection changes the behaviour of the post-MS evolution of a massive star, with an emphasis on the surface chemical composition and the pulsational properties of our models. In \citet{Saio2013a}, we showed that models of BSGs computed with the Schwarzschild criterion well reproduced the pulsational properties of BSGs only after a strong mass loss during the RSG phase (group 2). However, our models failed to reproduce the observed surface abundances of such stars. Using the Ledoux criterion for convection, the agreement between our results and the observed surface abundances is strongly improved, while the predicted pulsation periods are still in good agreement with the observed one.

The discussion on whether one has to adopt the Ledoux or the Schwarzschild criterion is a long lasting problem in stellar evolution and of course we cannot pretend to close the discussion here by concluding that Ledoux has to be definitely preferred. First, as usual, more observational data have to be collected (here the conclusions are based on only two stars). Second, we cannot be completely sure at the moment that the solution proposed here is unique, in the sense that some other combinations of overshoot, semi-convection or treatment of convective mixing may also provide a reasonable solution while keeping the Schwarzschild criterion. However, an interesting outcome of the present work is to illustrate through an original problem (i.e. how to make  the pulsational and surface abundances properties of BSGs in the group 2 compatible), the strong impact of the way the limit of the convective zones are computed. Another way to discriminate between both criteria is the ratio of red- to blue-supergiants. This ratio is well reproduced by our standard models \citep{Ekstrom2013a}. Using the Ledoux criterion increases the duration of the RSG stage by a factor of about 2, and changes only marginally the BSG duration. However, this ratio is highly sensitive to the mass-loss prescription, and it is yet not possible, on the basis of the two models presented here, to use it as a strong observational constraint. We are confident that the accumulations of observational constraints will narrow the set of possible solutions and will allow to chose the criterion that has to be adopted.

\section*{Acknowledgments}

The authors would like to warmly thank Arlette Noels who suggested us to investigate the impact of using the Ledoux criterion instead of the Schwarzschild one, as well as the anonymous referee for precious comments on this work. CG acknowledges support from EU-FP7-ERC-2012-St Grant 306901.

\bibliographystyle{aa}
\bibliography{MyBiblio}

\begin{thebibliography}{19}
\expandafter\ifx\csname natexlab\endcsname\relax\def\natexlab#1{#1}\fi

\bibitem[{{Bresolin} {et~al.}(2004){Bresolin}, {Pietrzy{\'n}ski}, {Gieren},
  {Kudritzki}, {Przybilla}, \& {Fouqu{\'e}}}]{Bresolin2004a}
{Bresolin}, F., {Pietrzy{\'n}ski}, G., {Gieren}, W., {et~al.} 2004, \apj, 600,
  182

\bibitem[{{Brott} {et~al.}(2011){Brott}, {de Mink}, {Cantiello}, {Langer}, {de
  Koter}, {Evans}, {Hunter}, {Trundle}, \& {Vink}}]{Brott2011a}
{Brott}, I., {de Mink}, S.~E., {Cantiello}, M., {et~al.} 2011, \aap, 530, A115

\bibitem[{{Ekstr{\"o}m} {et~al.}(2012){Ekstr{\"o}m}, {Georgy}, {Eggenberger},
  {Meynet}, {Mowlavi}, {Wyttenbach}, {Granada}, {Decressin}, {Hirschi},
  {Frischknecht}, {Charbonnel}, \& {Maeder}}]{Ekstrom2012a}
{Ekstr{\"o}m}, S., {Georgy}, C., {Eggenberger}, P., {et~al.} 2012, \aap, 537,
  A146

\bibitem[{{Ekstr{\"o}m} {et~al.}(2013){Ekstr{\"o}m}, {Georgy}, {Meynet},
  {Groh}, \& {Granada}}]{Ekstrom2013a}
{Ekstr{\"o}m}, S., {Georgy}, C., {Meynet}, G., {Groh}, J., \& {Granada}, A.
  2013, in EAS Publications Series, Vol.~60, EAS Publications Series, ed.
  P.~{Kervella}, T.~{Le Bertre}, \& G.~{Perrin}, 31--41

\bibitem[{{Georgy}(2012)}]{Georgy2012a}
{Georgy}, C. 2012, \aap, 538, L8

\bibitem[{{Georgy} {et~al.}(2013{\natexlab{a}}){Georgy}, {Ekstr{\"o}m},
  {Eggenberger}, {Meynet}, {Haemmerl{\'e}}, {Maeder}, {Granada}, {Groh},
  {Hirschi}, {Mowlavi}, {Yusof}, {Charbonnel}, {Decressin}, \&
  {Barblan}}]{Georgy2013b}
{Georgy}, C., {Ekstr{\"o}m}, S., {Eggenberger}, P., {et~al.}
  2013{\natexlab{a}}, \aap, 558, A103

\bibitem[{{Georgy} {et~al.}(2013{\natexlab{b}}){Georgy}, {Ekstr{\"o}m},
  {Granada}, {Meynet}, {Mowlavi}, {Eggenberger}, \& {Maeder}}]{Georgy2013a}
{Georgy}, C., {Ekstr{\"o}m}, S., {Granada}, A., {et~al.} 2013{\natexlab{b}},
  \aap, 553, A24

\bibitem[{{Georgy} {et~al.}(2012){Georgy}, {Ekstr{\"o}m}, {Meynet}, {Massey},
  {Levesque}, {Hirschi}, {Eggenberger}, \& {Maeder}}]{Georgy2012b}
{Georgy}, C., {Ekstr{\"o}m}, S., {Meynet}, G., {et~al.} 2012, \aap, 542, A29

\bibitem[{{Groh} {et~al.}(2013{\natexlab{a}}){Groh}, {Meynet}, \&
  {Ekstr{\"o}m}}]{Groh2013a}
{Groh}, J.~H., {Meynet}, G., \& {Ekstr{\"o}m}, S. 2013{\natexlab{a}}, \aap,
  550, L7

\bibitem[{{Groh} {et~al.}(2013{\natexlab{b}}){Groh}, {Meynet}, {Georgy}, \&
  {Ekstr{\"o}m}}]{Groh2013b}
{Groh}, J.~H., {Meynet}, G., {Georgy}, C., \& {Ekstr{\"o}m}, S.
  2013{\natexlab{b}}, \aap, 558, A131

\bibitem[{{Levesque} {et~al.}(2005){Levesque}, {Massey}, {Olsen}, {Plez},
  {Josselin}, {Maeder}, \& {Meynet}}]{Levesque2005a}
{Levesque}, E.~M., {Massey}, P., {Olsen}, K.~A.~G., {et~al.} 2005, \apj, 628,
  973

\bibitem[{{Meynet} {et~al.}(2013){Meynet}, {Ekstr\"om}, {Maeder},
  {Eggenberger}, {Saio}, {Chomienne}, \& {Haemmerl{\'e}}}]{Meynet2013a}
{Meynet}, G., {Ekstr\"om}, S., {Maeder}, A., {et~al.} 2013, in Lecture Notes in
  Physics, Berlin Springer Verlag, Vol. 865, Studying Stellar Rotation and
  Convection, ed. M.~{Goupil}, K.~{Belkacem}, C.~{Neiner}, F.~{Ligni{\`e}res},
  \& J.~J. {Green}, 3--642

\bibitem[{{Meynet} \& {Maeder}(2002)}]{Meynet2002a}
{Meynet}, G. \& {Maeder}, A. 2002, \aap, 390, 561

\bibitem[{{Przybilla} {et~al.}(2010){Przybilla}, {Firnstein}, {Nieva},
  {Meynet}, \& {Maeder}}]{Przybilla2010a}
{Przybilla}, N., {Firnstein}, M., {Nieva}, M.~F., {Meynet}, G., \& {Maeder}, A.
  2010, \aap, 517, A38+

\bibitem[{{Saio} {et~al.}(2013){Saio}, {Georgy}, \& {Meynet}}]{Saio2013a}
{Saio}, H., {Georgy}, C., \& {Meynet}, G. 2013, \mnras, 433, 1246

\bibitem[{{Smartt} {et~al.}(2009){Smartt}, {Eldridge}, {Crockett}, \&
  {Maund}}]{Smartt2009a}
{Smartt}, S.~J., {Eldridge}, J.~J., {Crockett}, R.~M., \& {Maund}, J.~R. 2009,
  \mnras, 395, 1409

\bibitem[{{van Loon} {et~al.}(2005){van Loon}, {Cioni}, {Zijlstra}, \&
  {Loup}}]{vanLoon2005a}
{van Loon}, J.~T., {Cioni}, M.-R.~L., {Zijlstra}, A.~A., \& {Loup}, C. 2005,
  \aap, 438, 273

\bibitem[{{Vanbeveren} {et~al.}(1998{\natexlab{a}}){Vanbeveren}, {De Donder},
  {van Bever}, {van Rensbergen}, \& {De Loore}}]{vanBeveren1998a}
{Vanbeveren}, D., {De Donder}, E., {van Bever}, J., {van Rensbergen}, W., \&
  {De Loore}, C. 1998{\natexlab{a}}, \na, 3, 443

\bibitem[{{Vanbeveren} {et~al.}(1998{\natexlab{b}}){Vanbeveren}, {De Loore}, \&
  {Van Rensbergen}}]{vanBeveren1998b}
{Vanbeveren}, D., {De Loore}, C., \& {Van Rensbergen}, W. 1998{\natexlab{b}},
  \aapr, 9, 63

\end{thebibliography}

\end{document}